\documentclass[aps,prc,superscriptaddress,floatfix,epsfig,preprint]{revtex4}
\usepackage{amsmath,amssymb,graphicx}

\begin{document}

\vspace{2.5cm}
\begin{center}
   {\LARGE \bf Electric charge fluctuations}
\end{center}
\begin{center}
 {\LARGE \bf in central Pb+Pb collisions }
\end{center}
\begin{center}
 {\LARGE \bf at 20, 30, 40, 80 and 158 {\it A}GeV}  
\end{center}

\vspace{1.5cm}
\begin{center}
   {\Large \bf The NA49 Collaboration}
\end{center}

\vspace{1.5cm}
\noindent
{\it 
Results are presented on event-by-event electric charge fluctuations in
central Pb+Pb collisions at 20, 30, 40, 80 and 158 $A$GeV. The observed
fluctuations are close to  those expected for a gas of pions
correlated by global charge conservation only. These fluctuations are
considerably larger than those calculated for an ideal gas of deconfined
quarks and gluons. The present measurements do not 
necessarily exclude reduced
fluctuations from a quark-gluon plasma because these might be masked by
contributions from resonance decays.
}

\newpage

\noindent
C.~Alt$^{9}$, T.~Anticic$^{21}$, B.~Baatar$^{8}$,D.~Barna$^{4}$,
J.~Bartke$^{6}$,  M.~Behler$^{13}$,
L.~Betev$^{9}$, H.~Bia{\l}\-kowska$^{19}$, A.~Billmeier$^{9}$,
C.~Blume$^{7}$
% ...

\vspace{0.5cm}
\noindent
% ...
$^{4}$KFKI Research Institute for Particle and Nuclear Physics, Budapest, Hungary.\\
% ...
$^{6}$Institute of Nuclear Physics, Cracow, Poland.\\
$^{7}$Gesellschaft f\"{u}r Schwerionenforschung (GSI), Darmstadt, Germany.\\
$^{8}$Joint Institute for Nuclear Research, Dubna, Russia.\\
$^{9}$Fachbereich Physik der Universit\"{a}t, Frankfurt, Germany.\\
% ...
$^{13}$Fachbereich Physik der Universit\"{a}t, Marburg, Germany.\\
% ...
$^{19}$Institute for Nuclear Studies, Warsaw, Poland.\\
% ...
$^{21}$Rudjer Boskovic Institute, Zagreb, Croatia.\\

$\;$

$\;$

$\;$



\newpage
\section{Introduction}

Ultra-relativistic heavy-ion collisions provide the opportunity 
to study the properties of strongly interacting matter. 
One of the predicted features of this matter, 
which one hopes to establish in heavy-ion collisions, 
is the occurrence of a phase transition between a purely
hadronic state and the quark-gluon plasma. 
Recently several results were reported 
\cite{Afanasiev:2002mx,Gazdzicki:2004ef} 
which suggest
that  this transition starts in central Pb+Pb collisions 
at energies around 30 {\it A}GeV \cite{Gazdzicki:1998vd,Gorenstein:2003cu}. 
The search for further signals of deconfinement is in progress 
and may provide additional support for such an interpretation.
Among them a suppression of event-by-event fluctuations of electric 
charge was predicted~\cite{Ko.1,As.1} 
as a consequence of deconfinement.
Estimates of the magnitude of the charge fluctuations 
indicate that  they are much smaller
in a quark-gluon plasma than in a hadron gas. 
Thus, naively, a decrease of the fluctuations 
is expected when the collision energy crosses the threshold
for the deconfinement phase transition. 
However, this  prediction is derived under the assumptions that 
the initial fluctuations survive hadronization 
and that their relaxation times in hadronic matter are significantly 
longer than the hadronic stage of the collision~\cite{Ko.1,As.1,Sh.1}.
First data on charge fluctuations in central heavy ion collisions
were published by PHENIX \cite{Adcox:2002mm} 
and STAR \cite{Adams:2003st}  at the BNL RHIC, and 
preliminary results at the CERN SPS were shown by 
NA49 \cite{Blume:2002mr}.  
The predicted large suppression of charge  fluctuations was not observed.
Results by  NA49 on 
transverse momentum 
and strangeness
fluctuations can be found in Refs. \cite{Appelshauser:1999ft,Anticic:2003fd} and
\cite{Afanasev:2000fu,Roland:2004pu}, respectively.

In this work final results on electric charge fluctuations 
in central Pb+Pb collisions at 20, 30, 40, 80 and 158 {\it A}GeV 
measured by NA49 at the CERN SPS are presented 
and discussed in view of their significance as a signal of deconfinement. 
The used measure of charge fluctuations 
$\Delta\Phi_{q}$ \cite{Za.1} is introduced 
in Sec. \ref{measure}. 
The experimental set-up is presented in Sec. \ref{exp} 
and the data sets as well as analysis method are described in Sec. \ref{ana}. 
Results are given in Sec. V and are discussed in Sec.
VI. The summary is given in Sec. VII.

\section{The measure of charge fluctuations}
\label{measure}

The magnitude of the measured charge fluctuations depends not only on 
the unit of electric charge carried by degrees of freedom of 
the system (hadrons or quarks and gluons), 
but depends also on trivial effects, 
which may obscure the physics of interest. 
The two most important of these effects 
are the fluctuations in the event multiplicity, 
caused mostly by the variation of the impact parameter, 
and changes in the mean multiplicity due to changes of the  
acceptance 
in which fluctuations are studied. 
In addition to the $\tilde{D}$  measure of charge fluctuations
\cite{Ko.1} several alternative measures such as $\nu_{+-,dyn}$, \cite{Star.1} and
$\Delta\Phi_q$ \cite{Za.1} were proposed to minimize the sensitivity to these
effects. In this analysis we use $\Delta \Phi_q$ which is constructed
from the well established measure $\Phi$ of event-by-event
fluctuations, defined as \cite{Ma.1}:

\begin{equation}\label{Phi}
\Phi = 
\sqrt{\langle Z^2 \rangle \over \langle N \rangle } -
\sqrt{\overline{z^2}} \; ,
\end{equation} 

where:
\begin{equation}\label{z}
z = x - \overline{x},\;\;\;\;\;\;\;\;
Z = \sum_{i=1}^{N}(x_i - \overline{x}) . 
\end{equation}

\noindent 
In these equations
$x$ denotes a single particle variable, $N$ is the number of  particles 
of the event
within the acceptance, and over-line and $\langle ... \rangle$ 
denote averaging over 
a single particle inclusive distribution and over events, respectively. 
By construction, for a system which is an independent sum of identical 
sources of particles the value of $\Phi$ is equal to the value of $\Phi$ 
for a single  source 
and does not depend on the number of superimposed 
sources  \cite{Ma.1,Ma.2}. 
In this analysis $x$ in Eqs. (\ref{z}) is taken to be
the electric charge $q$ and the measure is called
$\Phi_q$. 

For a scenario in which particles are correlated only by  
global charge conservation (GCC)   
the value of $\Phi_q$ is given by  
\cite{Za.1,Mrowczynski:2001mm}:
\begin{equation}\label{Phiq1}
\Phi_{q,\rm{GCC}}=\sqrt{1-P}-1 , 
\end{equation}
where
\begin{equation}\label{P}
P=\frac {\langle N_{ch}\rangle }{\langle N_{ch} \rangle _{tot}} 
\end{equation}
with $\langle N_{ch}\rangle$ and $\langle N_{ch} \rangle _{tot}$
being the mean charged multiplicity in the detector
acceptance and in full phase space (excluding spectator nucleons),
respectively. 
Strictly speaking Eq.~(3) holds for
vanishing net charge. However, as shown in \cite{Mrowczynski:2001mm}, Eq. (3)
serves as a  good approximation  also for realistic non-zero
values of the total net charge.

In order to remove the sensitivity to  GCC the measure $\Delta\Phi_{q}$ is 
defined as the difference:  
\begin{equation}\label{deltaPhi}
\Delta \Phi_q =\Phi_q-\Phi_{q,\rm{GCC}}\;.
\end{equation} 

\noindent By construction, the value of $\Delta \Phi_q$ is zero 
if the particles are correlated by global charge conservation only. 
It is negative in case of an additional correlation between positively 
and negatively charged particles, and it is positive if the positive 
and negative particles are anti-correlated \cite{Za.1}.

\section{Experimental Set-up}
\label{exp}

\begin{figure}[h]
  \begin{center}
    \includegraphics[height=5cm]{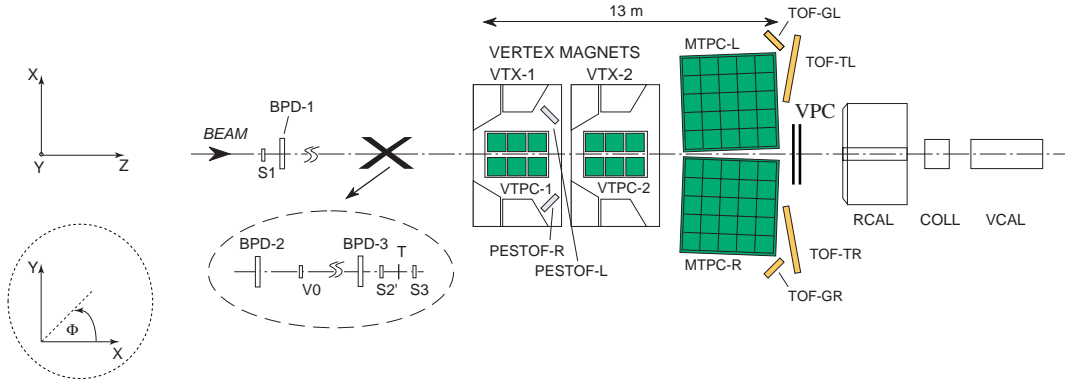}
    \caption{The experimental set-up of the NA49 experiment.}
    \label{SetUp}
  \end{center}
\end{figure}

The NA49 experimental set-up \cite{Inst} is shown in Fig. \ref{SetUp}. 
The main detectors 
of the experiment are four large-volume Time Projection Chambers (TPCs). 
Two of these, the Vertex TPCs (VTPC-1 and VTPC-2), are located in 
the magnetic field of two super-conducting dipole magnets. 
This allows separation of positively and negatively charged tracks 
and a measurement of the particle momenta. 
The nominal magnetic field is adjusted in proportion to the beam energy 
to ensure a good acceptance at all energies.  
The other two TPCs (MTPC-L and MTPC-R), positioned downstream of the magnets, 
are optimized for precise measurement of the ionization energy 
loss $dE/dx$ which is used for the determination of the particle masses. 
Additional information on the particle masses is provided by two 
Time-of-Flight (TOF) detector arrays which are placed behind the MTPCs. 
The centrality of the collisions is determined by a calorimeter (VCAL) 
which measures the energy of the projectile spectators. 
To cover only the spectator region the
geometrical acceptance
of the VCAL was adjusted for each beam energy 
by a proper setting of a collimator (COLL) \cite{Inst,Ap.98}.
The Beam Position Detectors (BPD-1, BPD-2 and BPD-3) are used to 
determine the $x$- and $y$-coordinate of the beam at the target. 
Alternatively the main vertex position is reconstructed as the
common intersection point of reconstructed tracks. 
A detailed description of the NA49 set-up and tracking software 
can be found in Ref.~\cite{Inst}. 

\section{Data Analysis}
\label{ana}
\subsection{Data}

At all energies 50K events were analyzed with a centrality of 7\%
of the inelastic cross-section except at 158 $A$GeV where the 10\%
most central events were selected.

\noindent To minimize the contributions of non-target collisions 
only events which satisfy the following  two selection 
criteria were used in the analysis.
Firstly, the reconstruction of the primary vertex position based 
on BPD and TPC data had to be successful. 
Secondly, the difference between vertex coordinates resulting 
from the BPD and TPC data 
should be 
smaller than $\pm 1$ mm in $x$- and $y$-coordinate 
and $\pm 5$ mm in $z$-coordinate. 

Several quality criteria were applied to  the particle tracks. 
All tracks should contain points measured in at least 
one of the Vertex TPCs and the number of all measured points, $n_P$, 
should be larger than 30. 
Only for these  particles charge and  momentum determination is
considered to be reliable.
To avoid double counting of particles only tracks with a measured number 
of points larger than 50\% of all geometrically possible points were accepted. 
The number of particles originating from weak decays and secondary 
interactions is reduced by only using  tracks for which the 
$x$- and $y$-position 
extrapolated to the $z$-coordinate of the target 
is close to the position of the 
interaction point ($|b_x| < 2$ cm for the $x$- and $|b_y| < 1$ 
cm for the $y$-coordinate).

Furthermore particles are required to lie in a well defined
  acceptance region in $y$, $p_T$ and $\phi$ 
($y$ 
is the rapidity in the center-of-mass system calculated assuming the pion mass, 
$p_T$ is the transverse momentum and $\phi$ denotes  the azimuthal angle). 
A well defined acceptance is essential for  comparison of 
the results with model predictions and with data from other experiments. 
The acceptance limits are parametrized by the function:
\begin{equation}
p_{T}(y,\phi)=\frac{1}{A(y)+(\frac{D(y)+\phi}{C(y)})^{6}}+B(y),
\end{equation}
where $A(y)$, $B(y)$, $C(y)$ and $D(y)$ are parameters depending 
on the rapidity and collision energy. 
The values of the parameters for positively charged tracks in the nominal
magnetic field ($B_y$ pointing upward) are given in Table 1. These
parameters also apply to negative tracks or to a reversed magnetic field
($B_y$ pointing downward) provided $\phi$ in Eq.~ (6) is replaced by $\phi' =
sign(\phi) (180 -|\phi|)$.

\begin{figure}[h]
  \begin{center}
    \includegraphics[height=5cm]{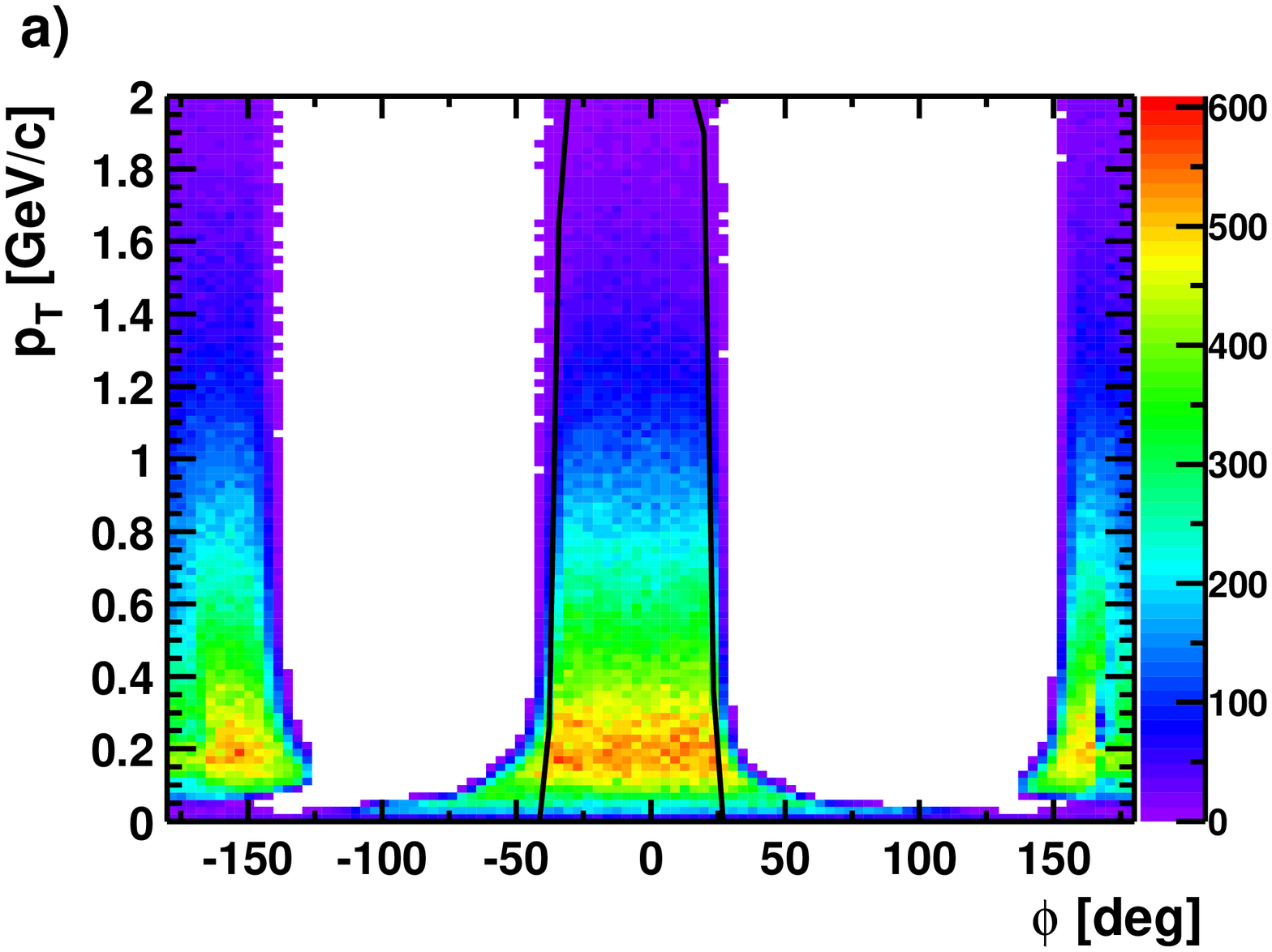}
    \includegraphics[height=5cm]{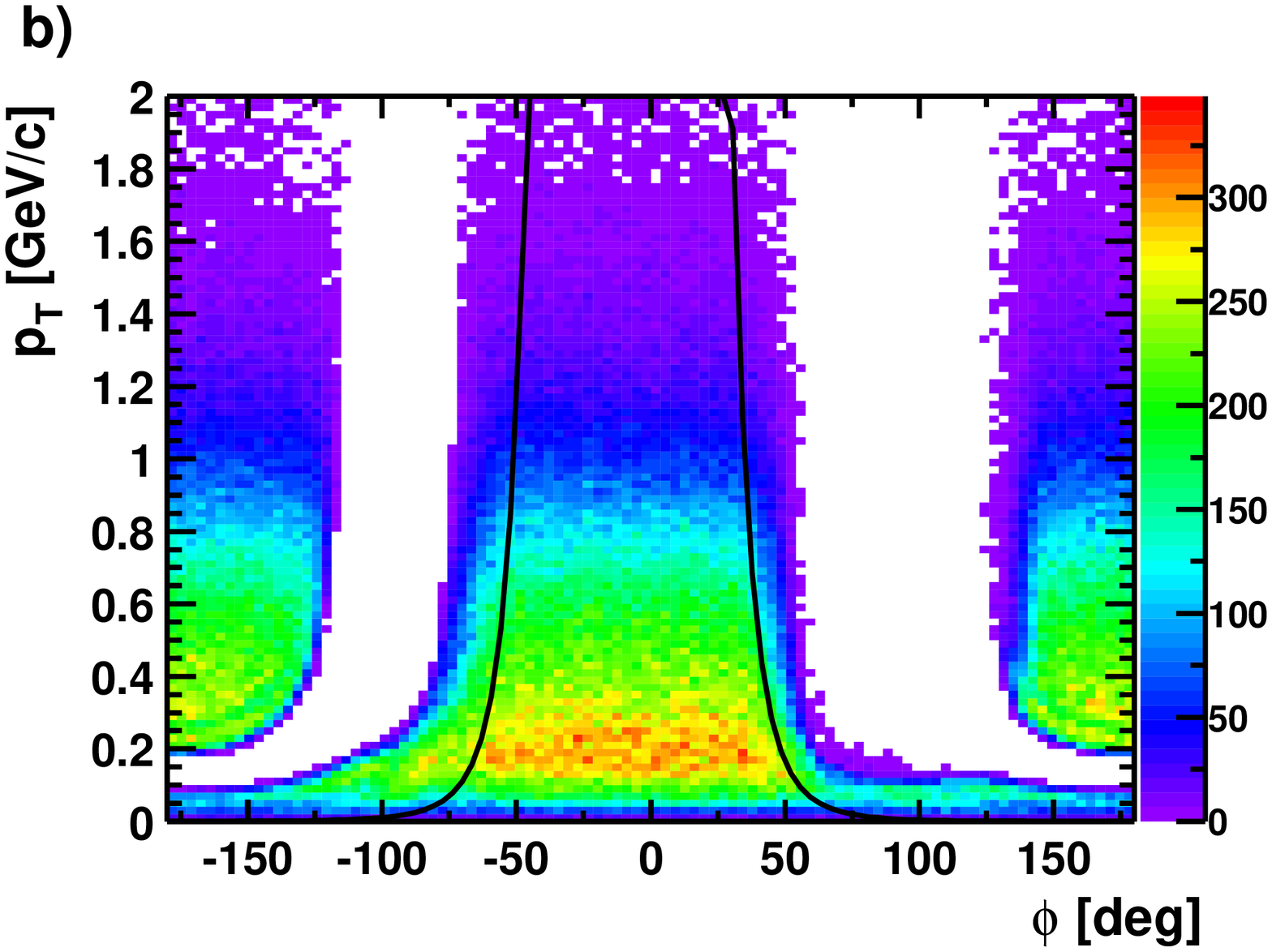}
    \includegraphics[height=5cm]{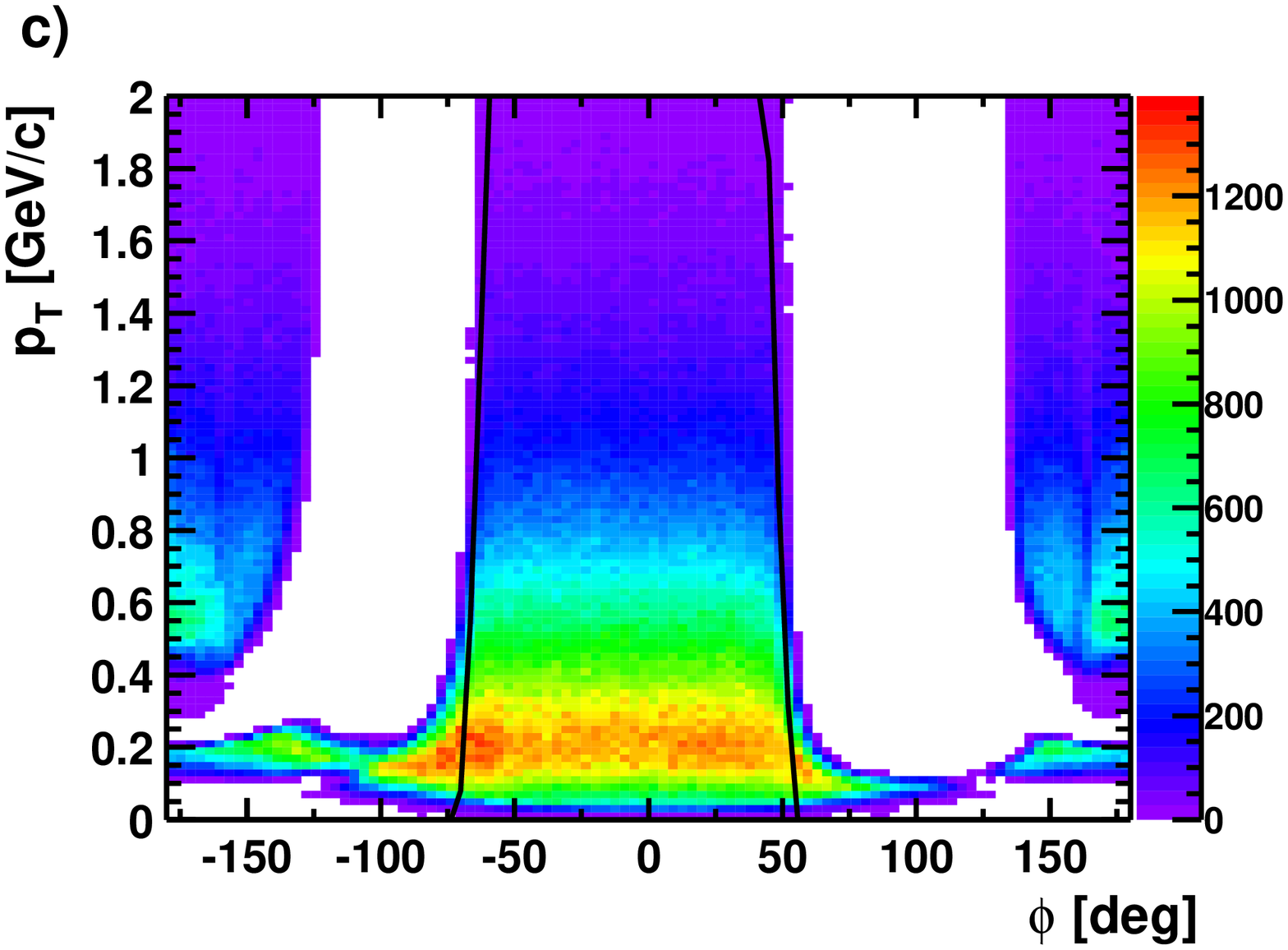}
    \includegraphics[height=5cm]{fig2b}
    \caption{
Distributions of registered particles in the  $p_{T}$-$\phi$-plane
for
 $-0.2<y<0$ (a) and $1.4<y<1.6$ (b) at 20~{\it A}GeV
and $-0.6<y<-0.4$ (c) and $1.4<y<1.6$ (d) at 158~{\it A}GeV.
The acceptance limits used in the analysis are shown by the solid lines.
}
    \label{accept}
  \end{center}
\end{figure}

As an example we show in   Fig.~\ref{accept} the acceptance in $p_{T}$, $\phi$  used 
for the analysis for $-0.2<y<0$ and $1.4<y<1.6$ at 20~{\it A}GeV
and $-0.6<y<-0.4$ and $1.4<y<1.6$ at 158~{\it A}GeV. 

\begin{table}[h]
\begin{center}
\caption{Values of the parameters $A$, $B$, $C$ and $D$ of 
the acceptance limits, Eq. 6, for different energies and rapidities.
The dimensions of the parameters are such that the use of the $\phi$
angle in degrees and the rapidity in the center-of mass system
results in the $p_T$ limit in GeV/c.}
\label{abcd}
\vspace*{0.5cm}
\begin{tabular}{c||c|c|c|c||c|c|c|c||c|c|c|c||c|c|c|c||c|c|c|c||} 
      &\multicolumn{4}{c}{20{\it A}GeV}&\multicolumn{4}{c}
{30{\it A}GeV}&\multicolumn{4}{c}{40{\it A}GeV}&\multicolumn{4}{c}
{80{\it A}GeV}&\multicolumn{4}{c}{158{\it A}GeV} \\ \hline
$y$    & A  & B   & C & D & A & B  & C & D & A & B  & C & D & A  & B  & C & D & A  & B  & C & D \\  \hline \hline
-0.5   &    &     &   &   &   &    &   &   & 0 & 0  &23 & 4 & 0  & 0  &35 &-10& 0  & -1 & 63&-8 \\ 
-0.3   &    &     &   &   & 0 & 0  &25 & -7& 0 & 0  &30 & 7 & 0  &0.07&40 &-10& 0  & 0  & 57&-10\\ 
-0.1   & 0  &  -1 &32 &-7 & 0 & 0  &31 & -8& 0 & 0  &38 &10 & 0  &0.07&46 &-10& 0  &0.09& 63&-13\\ 
0.1    & 0  &   0 &34 &-8 & 0 & 0  &40 & -8& 0 & 0  &43 & 8 & 0  &0.05&52 &-12& 0  &0.08& 67&-4 \\ 
0.3    & 0  &   0 &41 &-8 & 0 & 0  &44 & -8& 0 & 0  &46 & 7 & 0  & 0  &58 &-7 & 0  &0.08& 65&-3 \\
0.5    & 0  &-0.05&47 &-8 & 0 & 0  &46 & -7& 0 & 0  &40 & 0 & 0  & -1 &29 &-2 & 0  &0.05& 27& 0 \\
0.7    & 0  &-0.1 &50 &-7 & 0 & 0  &42 & 0 & 0 & 0  &22 & 0 & 0  &0.05&26 & 0 & 0  & 0  & 35& 0 \\
0.9    & 0  &-0.3 &53 &-3 & 0 & 0  &35 &-10& 0 & 0  &34 & 6 & 0  &0.08&35 & 0 & 0  &0.1 & 41& 0 \\
1.1    & 0  &-0.2 &38 &-10& 0 & 0  &39 &-13& 0 & 0  &46 &15 &0.3 &0.1 &67 &-27&0.34&0.43&109& 0 \\
1.3    & 0  &-0.1 &42 &-12& 0 & 0  &44 &-14& 0 & 0  &52 &15 &0.3 &0.3 &75 &-15&0.36&0.43&100& 0 \\
1.5    & 0  & 0   &43 &-8 & 0 & 0  &55 &-21& 0 &0.1 &58 &20 &0.3 &0.27&85 & 0 &0.55&0.4 &100& 0 \\
1.7    & 0  & 0   &51 &-18& 0 &0.08&62 & -2& 0 &0.08&72 & 0 &0.3 &0.18&75 & 0 &0.6 &0.4 & 88& 0 \\
1.9    & 0  & 0   &63 &-4 & 0 &0.08&67 & 0 & 0 &0.08&68 & 0 &0.45&0.15&70 & 0 &0.61&0.35& 73& 0 \\
2.1    & 0  & 0   &62 & 0 & 0 &0.05&61 & 0 & 0 &0.09&60 & 0 &0.5 &0.12&50 & 0 &0.73&0.34& 55& 0 \\
2.3    & 0  & 0   &57 & 0 &0.6&0.05&57 & 0 &0.5&0.08&50 & 0 &0.75&0.08&50 & 0 &1.7 &0.28& 60& 0 \\
2.5    &0.7 & 0   &54 & 0 &0.6& 0  &46 & 0 &0.6&0.05&40 & 0 &2.2 &0.08&50 & 0 &2.8 &0.25& 60& 0 \\
2.7    &0.7 & 0   &41 & 0 & 1 & 0  &33 & 0 &1.5&0.05&35 & 0 &3.2 &0.08&45 & 0 & 5  &0.2 & 57& 0 \\
2.9    &1.5 & 0   &30 & 0 &2.7& 0  &32 & 0 &   &    &   &   &4.5 &0.08&45 & 0 & 7  &0.15& 60& 0 \\
3.1    &    &     &   &   &   &    &   &   &   &    &   &   &5.5 & 0  &45 & 0 & 7  &0.1 & 70& 0 \\
\end{tabular}
\end{center}
\end{table}

\subsection{Analysis}

Charge fluctuations are studied as a function of the width of 
the rapidity interval $\Delta y$. 
These
rapidity intervals were centered around 
1.27, 1.07, 0.89, 0.89 and 0.89 for
the 20, 30, 40, 80 and 158~$A$GeV data, respectively. 
The measures  $\Phi_q$ and  $\Delta \Phi_q$ were
calculated for ten different rapidity intervals increasing in size from
$\Delta y = 0.3$ to $\Delta y = 3$ in equal steps
and will be plotted 
either versus $\Delta y$ or the corresponding ratio 
$\langle N_{ch} \rangle/\langle N_{ch} \rangle_{tot}$
The largest rapidity
interval contains approximately 90\% of all accepted particles. In Fig.~3
the rapidity distribution of the accepted particles at 158~$A$GeV is shown
together with the largest rapidity interval used in the
analysis.

\begin{figure}[h]
  \begin{center}
    \includegraphics[height=9cm]{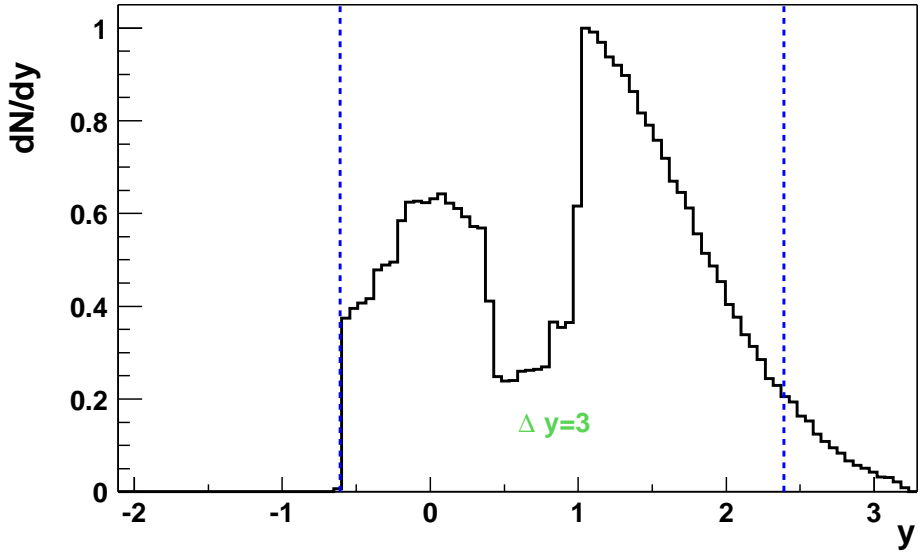}
    \caption{The rapidity distribution of accepted particles 
in central Pb+Pb collisions at 158 {\it A}GeV. 
The largest rapidity interval used for the analysis is indicated by
dashed lines.
} 
    \label{RapInt}
  \end{center}
\end{figure}

\noindent For each event the positively and negatively charged particles 
which fall into each rapidity interval and the corresponding $p_{T}$-$\phi$ 
acceptance are counted and using these numbers 
($N_+$ and $N_-$)
the values of $\Delta\Phi_{q}$ are
calculated.
The total charged particle multiplicity,
$\langle N_{ch} \rangle_{tot}$,
 was estimated for each energy
based on the NA49 measurements \cite{Afanasiev:2002mx,Gazdzicki:2004ef}.
\newline

\subsection{Errors}
The statistical error of $\Delta\Phi_{q}$ is calculated by dividing 
the whole sample of events into 10 subsamples and calculating 
$\Delta\Phi_{q}$ for each subsample separately. 
The dispersion  of the obtained $\Delta\Phi_{q}$ values divided by 
$\sqrt{9}$ has been taken as the statistical error.
The systematic errors of $\Delta\Phi_{q}$ are 
estimated by varying track quality cuts: 
The values of $\Delta\Phi_{q}$ are calculated for two additional sets of cuts, 
more ($n_P = 35, |b_x| < 0.75$ cm and $|b_y| < 0.5$ cm)
and less ($n_P = 30, |b_x| < 4.5$ cm and $|b_y| < 2.5$ cm)
restrictive in comparison to the standard cuts. 
The accepted particle multiplicity decreases by about 25\%
when changing from less to more restrictive cuts.
The difference of these two $\Delta\Phi_{q}$ 
values is considered as the systematic error. 
Since the statistical errors are much smaller  
only the systematic errors are shown in the figures.

\section{Results}
\label{results}

\begin{figure}[h]
  \begin{center}
    \includegraphics[height=9cm]{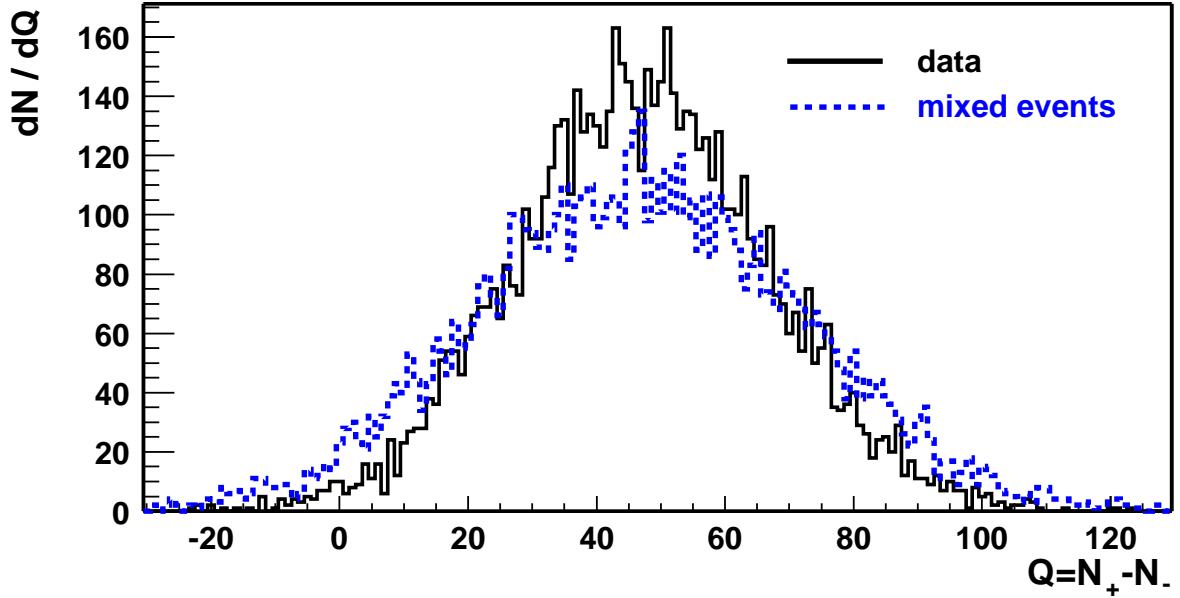}
    \caption{The distribution of the net-charge for central Pb+Pb collisions 
at 158 {\it A}GeV (solid line) and the corresponding distribution obtained 
for mixed events (dotted line) in the maximum rapidity interval
$\Delta y = 3$.}
    \label{Qdist}
  \end{center}
\end{figure}

A simple measure of charge fluctuations is the width of the 
distribution of net-charge 
$Q=N_{+}-N_{-}$ in the events. 
As an example the distribution for central Pb+Pb collisions 
at 158 {\it A}GeV is shown in Fig. \ref{Qdist}. 
This distribution is compared to the net-charge distribution obtained 
from mixed events (dashed line in Fig. \ref{Qdist}) 
constructed by randomly selecting particles from different events
according to the multiplicity distribution measured for the data.
The net-charge distribution from mixed
events is significantly broader 
than the net-charge distribution obtained from real events. 

\begin{figure}[h]
  \begin{center}
    \includegraphics[height=9cm]{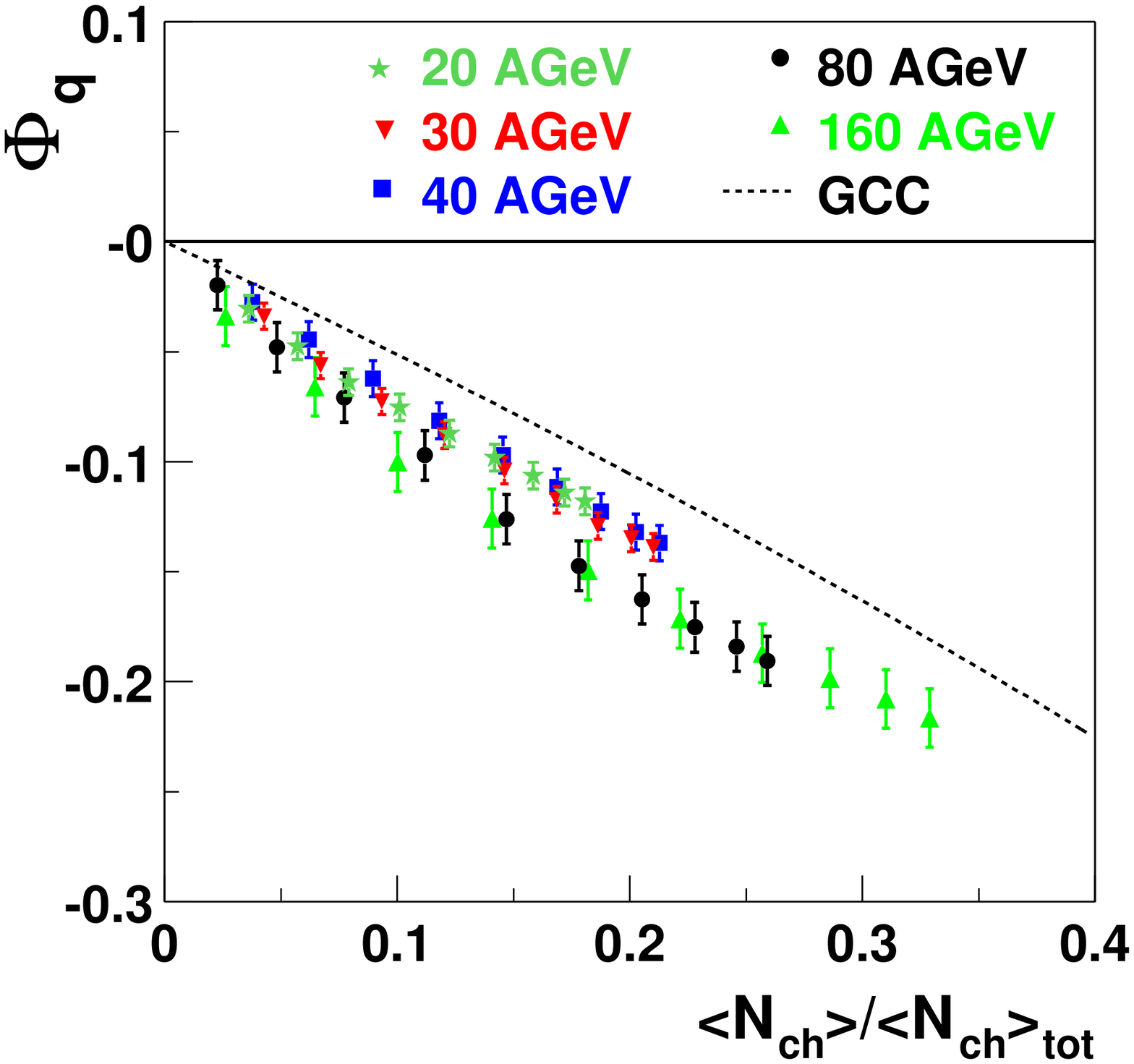}
    \caption{The dependence of the measure $\Phi_q$ on the fraction
of accepted particles for central Pb+Pb collisions at 20-158~{\it A}GeV. 
Note that experimental points for a given energy are correlated as 
the data used for a given rapidity interval
 are included in the broader intervals. 
The dashed line shows the dependence expected for the case when the only source
of particle correlations is the global charge conservation, 
Eq.~3.
}
    \label{nch}
  \end{center}
\end{figure}

The main source of the observed difference is charge conservation 
which correlates positively 
and negatively charged particles in the real, but not in the mixed events. 
This is demonstrated in Fig.~\ref{nch} where the $\Phi_q$ values
are plotted as a function of the fraction of accepted 
particles $\langle N_{ch} \rangle/\langle N_{ch} \rangle_{tot}$ for central
Pb+Pb collisions at 20-158~{\it A}GeV.
The main trend observed in the data, a monotonic decrease with 
increasing fraction of accepted particles, is approximately reproduced
by introducing global charge conservation as the only source of
particle correlations (Eq.~3 shown by the dashed line in Fig.~\ref{nch}). 

\begin{figure}[h]
  \begin{center}
    \includegraphics[height=9cm]{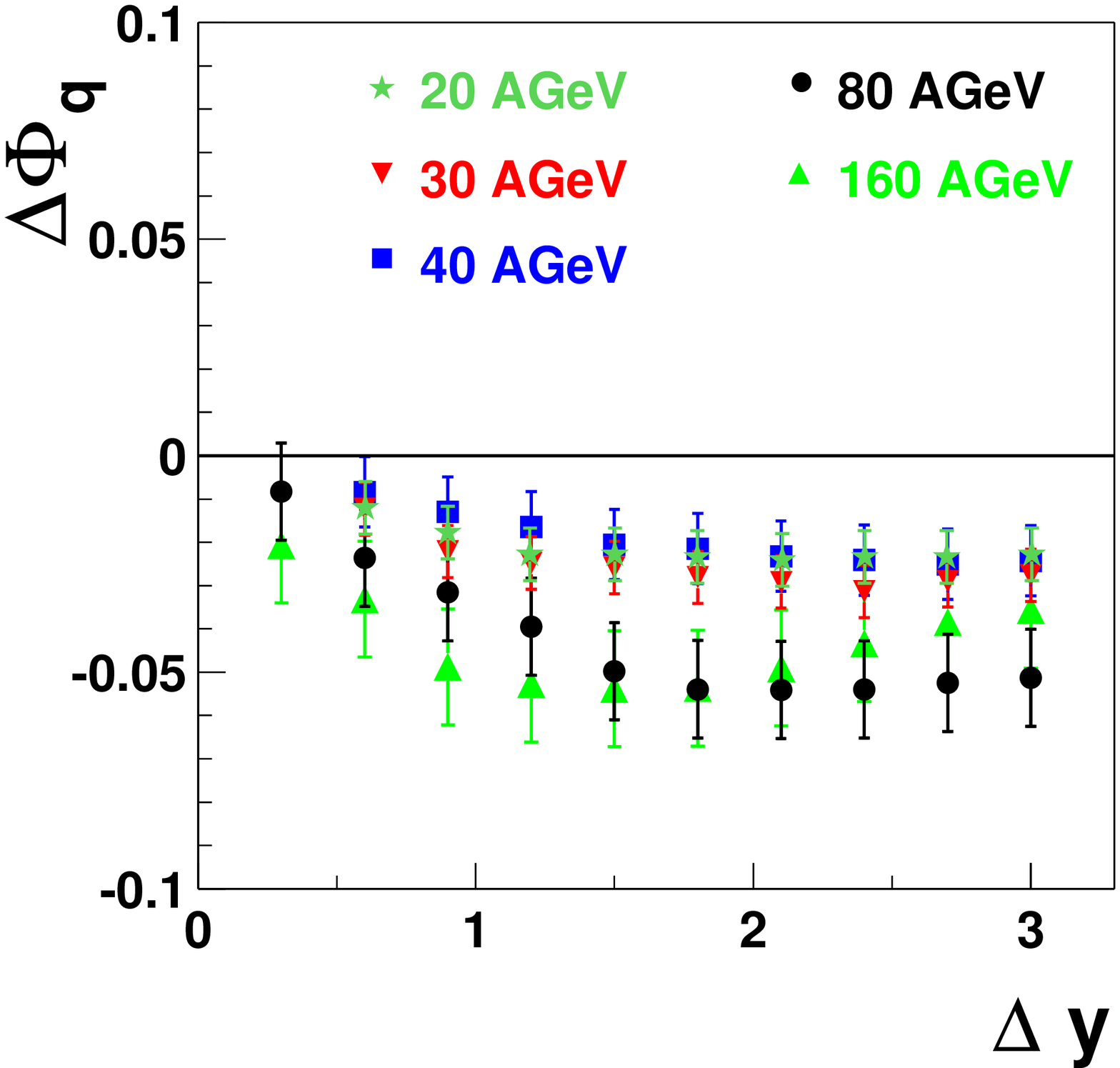}
    \caption{The dependence of $\Delta\Phi_{q}$ 
on the width of the rapidity interval $\Delta y$ for central Pb+Pb collisions 
at 20, 30, 40, 80 and 158 {\it A}GeV.
Note that experimental points for a given energy are correlated as 
the data used for a given rapidity interval are included in the broader intervals.} 
    \label{AllEnergies}
  \end{center}
\end{figure}

\begin{figure}[h]
  \begin{center}
    \includegraphics[height=8cm]{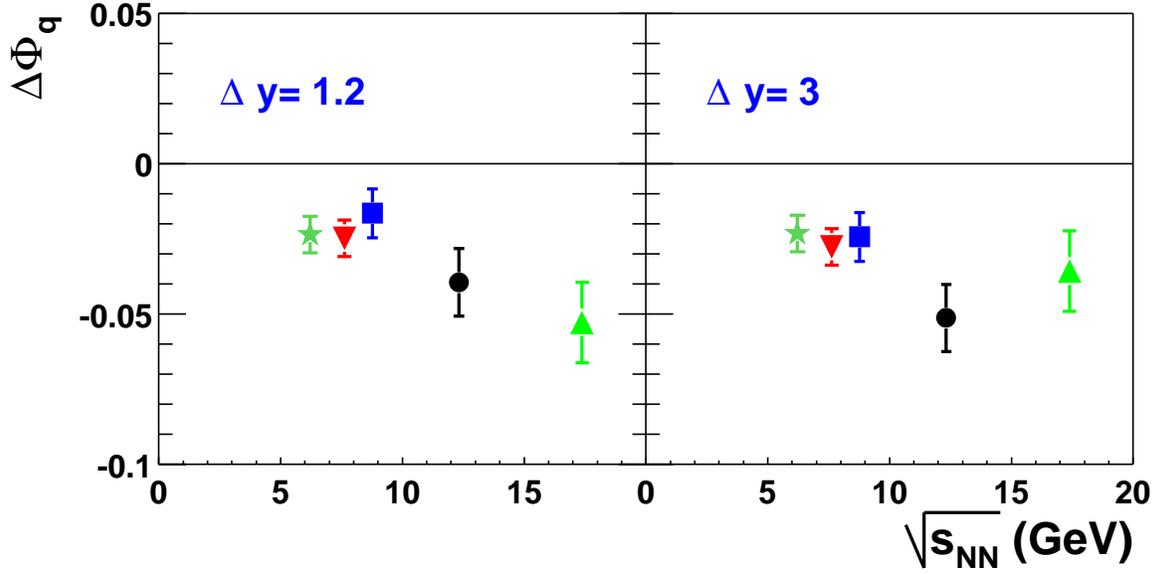}
    \caption{The energy dependence of $\Delta\Phi_{q}$ 
measured in central Pb+Pb collisions  for 
a narrow rapidity interval $\Delta y = 1.2$
(left) and a broad rapidity interval $\Delta y = 3$ (right).}
    \label{AllBig}
  \end{center}
\end{figure}

By construction, the previously introduced
measure $\Delta\Phi_q$ is
insensitive to the correlations due to
global charge conservation  
(see Sec. \ref{measure}).
The dependence of $\Delta\Phi_q$ on 
the width of the rapidity interval $\Delta y$
is shown in Fig. \ref{AllEnergies} 
for central Pb+Pb collisions at 20, 30, 40, 80 
and 158 {\it A}GeV. 
The study of charge fluctuations as a function of $\Delta y$
was suggested in the original proposal
\cite{Ko.1,As.1}. 
The measured values of $\Delta\Phi_q$ vary between $0$ and  $-0.05$. 
They are significantly larger than the values expected for QGP fluctuations 
($-0.5<\Delta\Phi_q < -0.15$ \cite{Za.1,Za.2}). 
The energy dependences of $\Delta\Phi_q$ for the largest rapidity interval 
($\Delta y = 3$) and for the rapidity interval  
$\Delta y = 1.2$ are shown in  
Fig.~\ref{AllBig}. 
A weak decrease of $\Delta\Phi_q$ with increasing energy is
suggested by the data.
The numerical values of $\Delta\Phi_q$ for $\Delta y = 1.2$ and $\Delta y = 3$ 
are given in Table~\ref{Results}.
Note that the fraction of the accepted tracks for a fixed 
rapidity interval $\Delta y$ changes with collision energy,
but this alone should not affect $\Delta\Phi_q$ provided
the correlation length  is smaller than the
acceptance interval in $y$.

\begin{table}[h]
\begin{center}
\caption{
The values of 
$\Delta\Phi_q$ for $\Delta y= 3.00$ and for 
$\Delta y= 1.2$ in central Pb+Pb collisions 
at 20, 30, 40, 80 and 158 {\it A}GeV. The first error is systematic,
 the second statistical.}
\label{Results}
\vspace*{0.5cm}
\begin{tabular}{c|c|c|c} 
{\bf $\Delta\Phi_q$}    & {\bf 20 {\it A}GeV}   &{\bf 30 {\it A}GeV}    & {\bf 40 {\it A}GeV}  \\ \hline \hline
$\Delta y= 3 $  & -0.023$\pm$0.006$\pm$0.0001 & -0.028$\pm$0.0003$\pm$0.002 & -0.024$\pm$0.008$\pm$0.0005\\  
$\Delta y= 1.2$ & -0.023$\pm$ 0.006$\pm$ 0.0001& -0.025$\pm$0.0002$\pm$0.016 & -0.016$\pm$0.008$\pm$0.0003\\  
\end{tabular}
\begin{tabular}{c|c|c} 
{\bf $\Delta\Phi_q$}    & {\bf 80 {\it A}GeV}    & {\bf 160 {\it A}GeV}\\ \hline \hline
$\Delta y= 3 $  & -0.051$\pm$0.011$\pm$0.0002& -0.036$\pm$0.013$\pm$0.0003\\  
$\Delta y= 1.2$ & -0.040$\pm$0.011$\pm$0.0003& -0.053$\pm$0.013$\pm$0.0004\\  
\end{tabular}
\end{center}
\end{table}

%

\section{Discussion}

\begin{figure}[h]
  \begin{center}
    \includegraphics[height=9cm]{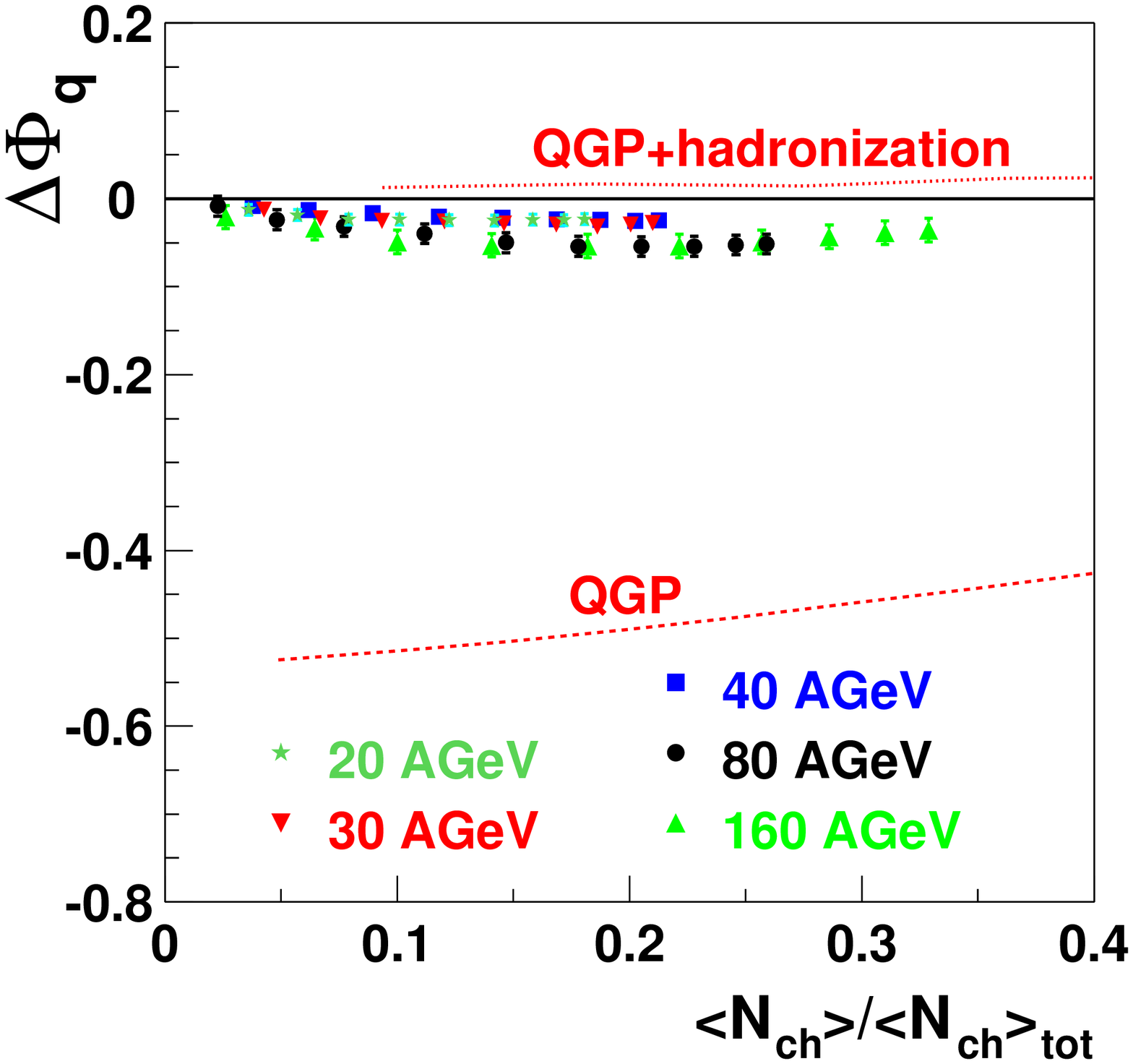}
    \caption{The dependence of $\Delta\Phi_q$ on the fraction of accepted particles
in central Pb+Pb collisions at 20-158 {\it A}GeV.
The prediction for the ideal QGP is indicated by the dashed curve (QGP),
whereas the prediction for the QGP including hadronization and resonance decay
is shown by the dotted curve (QGP+hadronization). }
    \label{QGP}
  \end{center}
\end{figure}

The study of charge fluctuations in A+A collisions was motivated
by the hypothesis that they may be sensitive to the creation of
the Quark Gluon Plasma at the early stage of the collisions.
To quantify the expected effect 
a simple QGP model was proposed in \cite{Ko.1}. 
In this model quarks and gluons are assumed to be in local equilibrium. 
Assuming entropy and net charge conservation during the evolution from 
the QGP to the final hadron state in each rapidity interval 
the number $N$ of pions  and their net charge is calculated. 
The number of charged pions is taken to be $N_{ch}=\frac{2}{3}\cdot N$ 
based on isospin symmetry. 
Using this model it was shown that the electric charge fluctuations 
are significantly smaller in the case of QGP creation than in
the case of formation of confined matter at the early stage of
the collisions
(see Fig.~\ref{QGP})~\cite{Ko.1,Za.1}. 

\noindent However, this model is not complete. 
A large fraction of pions originates from decays of resonances \cite{Be.1}. 
This effect is expected to lead to a distortion of the charge fluctuations 
established after hadronization. 
To quantify this effect the model was extended as follows.
From the total number of produced final state pions 
the entropy of the system is calculated. 
This entropy is attributed to the early stage QGP, treated as 
an ideal gas of massless quarks and gluons. 
Bose-Einstein- and Fermi-Dirac-statistics are used to calculate 
equilibrium numbers of quarks and gluons. 
The rapidity distribution of these partons is centered at $y=0$ 
and  is assumed to be of Gaussian shape with $\sigma = 0.8$. 
For the calculations of charge fluctuations the rapidity interval $-3<y<3$ is 
divided into several (10 and 20) bins and in each bin the entropy  
and the net-charge of the contained partons is calculated. 
The resulting values of $\Delta\Phi_q$ at the QGP level
are shown by the dashed line in Fig. 8.
In the next step
the QGP entropy is attributed to an ideal gas of $\rho$ mesons. 
The numbers of $\rho^{+}$, $\rho^{-}$ and $\rho^{o}$ mesons 
in each bin are calculated assuming that $\frac{1}{3}$ of all $\rho$ mesons are neutral. 
Furthermore, all $\rho$ mesons are assumed to decay  into two pions. 
The rapidity distribution of the pions is divided into 20 bins and 
the $\Delta\Phi_q$ is calculated from 
the number of positively and negatively charged pions in each bin.
The results of this model are shown in Fig.~\ref{QGP} 
by the dotted curve.
Note that by construction $\Delta\Phi_q = 0$ for the full
acceptance.

As expected, the decays of resonances strongly 
modify the initial QGP fluctuations. 
The value of $\Delta\Phi_q$ increases from values between $-0.4$ and $-0.5$ 
(the lower line in  Fig. \ref{QGP}) to values  close to zero
(the upper line in  Fig. \ref{QGP}), the  value
characteristic for 
a gas of pions correlated 
by global charge conservation only. 
The model demonstrates that the distribution of charged particles 
in the detector acceptance
is strongly distorted by the decay of intermediate resonance states. 
This may explain why the measurements do not show the suppression of 
the charge fluctuations naively expected in the case of QGP creation.  

The influence of resonance decays on charge fluctuations depends on 
the size of the  rapidity interval $\Delta y$, in which fluctuations are
calculated. 
If  $\Delta y$ is much bigger than the typical 
distance in rapidity  of the daughter particles, 
the charge within the interval will not be changed by the decay 
and therefore the charge fluctuations should not be affected. 
On the other hand, if  $\Delta y$ is small, 
a large fraction of daughter particles will leave the interval 
and the initial net-charge will be significantly changed. 
The mean rapidity difference of two pions originating from decays of 
a $\rho (770)$ meson is approximately 1 unit of rapidity. 
Therefore in order to minimize the decay effect the rapidity 
interval should be much larger than 1. 
However, this constraint is difficult to fulfill 
at SPS and lower energies because 
the rapidity distribution of all produced particles 
is not much broader than 1. 
This explains an approximately constant value of $\Delta\Phi_q$
calculated  within the QGP+hadronization model as seen in Fig.~\ref{QGP}.
A rapidity interval which is large enough to be unaffected by the influence of 
resonance decays would contain almost all particles produced in a collision. 
The net-charge in this interval would then reflect 
the number of participant protons 
and the  fluctuations would be determined  by fluctuations of 
the collision centrality and not the particle production 
mechanism.
Thus at SPS energies the measured charge fluctuations are 
not sensitive to the initial QGP fluctuations. 
At very high energies (when the rapidity distribution of 
produced particles is significantly broader than 1) the charge 
fluctuations may be a valid signature of QGP creation.

\section{Summary}
\label{summary}

Results on event-by-event charge fluctuations in central Pb+Pb collisions 
at 20, 30, 40, 80 and 158 {\it A}GeV are presented in terms 
of the $\Delta\Phi_q$ measure. \newline 
The measured $\Delta\Phi_q$  values are close to zero, as  
expected for a gas of pions correlated only by global charge fluctuations. 
This value is significantly higher than  that predicted  for 
the creation of a QGP and hadronization into pions with
  local conservation of entropy and net-charge.
A model which incorporates intermediate resonances
 is described in this paper. 
Its results show that the decay of $\rho$ mesons may easily increase 
the initial QGP charge fluctuations to $\Delta\Phi_q\approx 0$  
thereby completely masking a possible  QGP signal at SPS energies.  \newline
The slightly negative value of $\Delta\Phi_q$ indicates correlations 
between positively and negatively charged particles beyond those from 
global charge conservation. 
The origin of these additional correlations may be the final state Coulomb 
interactions or quantum-statistical effects.

\vspace{0.5cm}
\noindent
Acknowledgments: This work was supported by the US Department of Energy
Grant DE-FG03-97ER41020/A000,
the Bundesministerium fur Bildung und Forschung, Germany, 
the Polish State Committee for Scientific Research 
(2 P03B 130 23, SPB/CERN/P-03/Dz 446/2002-2004, 2 P03B 02418, 2 P03B 04123), 
the Hungarian Scientific Research Foundation (T032648, T032293, T043514),
the Hungarian National Science Foundation, OTKA, (F034707),
the Polish-German Foundation, 
and the Korea Research Foundation Grant (KRF-2003-070-C00015).

\end{document}